\documentclass[journal,twocolumn,comsoc]{IEEEtran}

\usepackage{color}
\usepackage{graphicx}
\graphicspath{{fig/}}
\DeclareGraphicsExtensions{.pdf,.eps,.jpeg,.png}
\usepackage{epstopdf}
\usepackage{color}
\usepackage{amsmath}
\usepackage{amssymb}
\usepackage[caption=false,subrefformat=parens,labelformat=parens,font=footnotesize]{subfig}

\usepackage{multirow}
\usepackage{algpseudocode}

\usepackage{algorithm}
\usepackage{array}

\usepackage{pgfplots}
\pgfplotsset{compat=newest}
\pgfplotsset{
    legend image with text/.style={
        legend image code/.code={%
            \node[anchor=center] at (0.3cm,0cm) {#1};
        }
    },
}
\pgfplotsset{ignore legend/.style={every axis legend/.code={\renewcommand\addlegendentry[2][]{}}}}

\usetikzlibrary{quotes,graphs,plotmarks}

\usepackage{grffile}
\pgfplotsset{plot coordinates/math parser=false}
\newlength\figureheight
\newlength\figurewidth
\newlength\subfloatwidth
\DeclareMathOperator{\E}{E}
\DeclareMathOperator{\var}{var}
\DeclareMathOperator{\cov}{cov}

\DeclareMathOperator{\re}{Re}
\DeclareMathOperator*{\argmax}{arg\,max}

\begin{document}
    \title{Channel Access Method Classification For Cognitive Radio Applications}
     \author{Mihir~Laghate,~\IEEEmembership{Student Member,~IEEE,}
             Paulo~Urriza,
            and~Danijela~Cabric,~\IEEEmembership{Senior~Member,~IEEE}%
\thanks{Mihir Laghate is with Qualcomm
    Technologies, San Diego, CA 92121, Paulo Urriza is with Marvell
    Semiconductor, Santa Clara, CA 95054, and Danijela Cabric is with the University of California, Los Angeles, CA
    90095. This work was done while Mihir and Paulo were PhD students at UCLA.\protect\\
    E-mail: mvlaghate@ucla.edu, pmurriza@ucla.edu, danijela@ee.ucla.edu.}%
    }
    \maketitle

    \begin{abstract}
        
        Motivated by improved detection and prediction of temporal holes, we propose a two stage algorithm to classify the channel access method used by a primary network.
        The first stage extends an existing fourth-order cumulant-based modulation classifier to distinguish between TDMA, OFDMA, and CDMA. The second stage proposes a novel collision detector using the sample variance of the same cumulant to detect contention-based channel access methods.
        Our proposed method is blind and independent of the received SNR. Simulations show that our classification of TDMA, OFDMA, and CDMA is robust to network load while detection of contention outperforms existing methods.

    \end{abstract}
    
    \begin{IEEEkeywords}
        Automatic signal classification, channel access method, collision detection. 
    \end{IEEEkeywords}

    \section{Introduction}
    \label{sec:intro}
    \IEEEPARstart{C}{ognitive} radios (CRs) require radio scene analysis in order to achieve a more efficient utilization of the scarce radio spectrum~\cite{haykin_cognitive_2005}. Various works that dealt with the radio-scene analysis problem have mainly focused on the binary hypothesis problem of detecting the presence or absence of spectrum opportunities, i.e., spectrum holes, through various spectrum sensing methods.
    Various methods proposed for quickly detecting spectrum holes and scheduling CR transmissions efficiently require
    knowing the medium access protocol used by the primary user (PU) network. In particular, schemes have been
proposed if the PU network uses a time slotted method~\cite{chen_joint_2008, wellisch_spectrum_2012}, orthogonal frequency
division multiple access (OFDMA)~\cite{tu_spectrum_2009, bernardo_selforganized_2009}, code division multiple access
(CDMA)~\cite{li_cognitive_2010,khoshkholgh_adaptive_2008}, or contention-based
methods~\cite{liu_channel_2014,suzuki_intelligent_2011,cheng_decentralized_2016}. Hence,
we are motivated to propose a method to classify the channel access method used by the PU network.
    \subsection{Related Work}

    There has been little work on identifying the channel access method used by the primary network. Existing works fall
    into two categories: identification of particular standards or identification of class of standards. First, work
    such as \cite{dibenedetto_automatic_2010} and \cite{hachemani_new_2007} aim to determine the specific standard used
    based on detailed knowledge of PHY/MAC characteristics such as packet structure and preamble format.
    \cite{rajab_energy_2015} uses supervised learning of the frame lengths and inter-arrival times to
    distinguish between members of the 802.11 family. The second category, which we will be focusing on, identify the
    class of standards, i.e., OFDMA for 802.11n and CDMA for CDMA2000. The only existing work that we know of in this
    category is \cite{hu_mac_2014}. The authors of \cite{hu_mac_2014} employ a Support Vector Machine (SVM) approach to
    classify between TDMA, carrier sense multiple access with collision avoidance (CSMA/CA), slotted ALOHA, and pure
    ALOHA networks. However, such a supervised learning approach has limitations in unknown fading channels due to the
    lack of labeled data. Further, CDMA and OFDMA systems are not addressed in~\cite{hu_mac_2014}. 

    \subsection{Contributions}
    
    The two key contributions of this work are as follows. First, we extend an existing fourth-order cumulant-based
    modulation type classifier~\cite{swami_hierarchical_2000} to distinguish between TDMA, CDMA, and OFDMA. Second, we
    propose a novel method for detecting collusins using the sample variance of the same cumulant and thus, detect
    contention-based channel access methods.

    The rest of the paper is organized as follows. The system model and notation is described in Section~\ref{sec:ProblemFormulation}. Our proposed method is described in Section~\ref{sec:ProposedApproach}. Simulation results and comparison with existing work is provided in Section~\ref{sec:Results}. Finally, the paper is concluded in Section~\ref{sec:Conclusion}.

    \section{System Model}
    \label{sec:ProblemFormulation}
    
    We consider a system consisting of a single sensing node receiving signals from a network of $N_\text{total}$ PUs communicating amongst themselves. Let $\mathcal{U}$ be the index set of the PUs. We model the signal transmitted by the $i$th PU as 
    \begin{equation}
        \label{eq:pu_signal_model}
        x_i(n) = a_i(n)s_i(n)
    \end{equation}
    where $n$ is the time index, $a_i(n) = 1$ if the $i$th PU is transmitting at time $n$ and 0 otherwise, and $s_i(n) \in \mathbb{C}$ is the signal transmitted by the PU, if active. Both the activity $a_i(n)$ and the signal $s_i(n)$ depend on the channel access method used by the PU network. %
    
    For both TDMA and contention-based channel access methods, we assume that $s_i(n)$ is a single carrier signal with linear memory-less modulation, such as QAM. 
    TDMA enforces orthogonality in time, i.e., $\sum_{i = 1}^{N_{\text{total}}} a_i(n) \leq 1$, while contention-based schemes do not, i.e., $\sum_{i=1}^{N_\text{total}} a_i(n) \in \mathbb{N} \cup \{0\}$.
    
    For OFDMA, $s_i(n)$ is an OFDM modulated signal with $N_{sc}$ subcarriers and a cyclic prefix of length $N_p$. Of these $N_{sc}$ subcarriers, subset $\mathcal{S}_i$ are assigned to the $i$th PU.
    
    For CDMA, the $i$th PU's data stream $d_i(n)$ is spread using its code $c_i(n)$ of length $L_c$:
    \begin{equation}
        \label{eq:cdma_sig_model}
        s_i(n) = c_i\left( n\!\!\! \mod L_{c}\right)d_i(n).
    \end{equation}
    
    For both OFDMA and CDMA, we assume that all PUs are transmitting simultaneously, i.e., $a_i(n) = 1$. Though this appears to be a strong assumption, squelching the received signal and normalization of the test statistic ensure that it does not impact the performance of our algorithm.
    
    We denote the tuple of channel access method and modulation type used by $\mathcal{M}$. Non-contention based channel 
    access methods covered by this work are listed in Table~\ref{tab:list_classes} and collected in the set $\mathfrak{M}$.

    The packet arrival rate at the PUs determines the rate of collisions in a contention-based scheme. For simplicity, we assume the following conditions: 1) all packets have the same length, 2) a packet is generated from each PU according to a Poisson process with rate $\lambda_i$, 3) packets that collide are not retransmitted. As a result, the aggregate messages to the channel will also be a Poisson arrival process with parameter $G=\sum_{i=1}^N \lambda_i$. This is referred to as offered load. %

    The downconverted signal received at the sensing node is
    \begin{equation}
        r\left(n\right)=\sum_{i=1}^{N_\text{total}}h_{i}\left(n\right)x_{i}(n)+\nu\left(n\right)
        \label{eq:sig_model}
    \end{equation}
    where $\nu(n)\sim\mathcal{CN}(0,\sigma_\nu^2)$ is white Gaussian noise with variance $\sigma_\nu^2$ and $h_i(n)$ includes the PU's transmit power, the fading channel from the PU to the sensing node, and path loss with exponent $\gamma$. Let $y(n)$ be the squelched received signal.

    \section{Identifying Channel Access Method}
    \label{sec:ProposedApproach}
    
    TDMA, OFDMA, and CDMA have distinguishable features due to their signal structure viz., modulation used in TDMA, the
    effect of the inverse fourier transform in OFDMA, and the pseudonoise sequence used by users in CDMA.
    Contention-based methods are distinguished by the probability of collisions between users. We now show that these
    properties can be distinguished using the sample mean and sample variance of the normalized fourth-order cumulant of
    the squelched signal.

    \subsection{Normalized 4th-Order Cumulant $C_{42}$ and its Properties}
    \label{sec:CumulantIntro}
    
    Cumulants of multiple random variables are particular polynomial combinations of their joint higher order moments \cite{brillinger_time_2001}. 
    We define $k$th-order cumulants $C_{ki}$, for $i \in \{0, \ldots, k\}$, of the zero mean complex signal $y(n)$ as the cumulant of $k-i$ repeated copies of $y(n)$ and $i$ repeated copies of $y^*(n)$. Also, let $M_{ki} \triangleq E[y^{k-i}(n)(y^*(n))^i].$

    Our proposed test statistic requires estimating the cumulants of frames of $J$ samples each. The unbiased maximum likelihood estimators for the cumulants of the $f$th frame are denoted by $\hat{C}_{ki}(f)$ and are computed by formulae derived in \cite{swami_hierarchical_2000}.
    Similar to \cite{swami_hierarchical_2000}, the cumulant estimates are normalized by the signal power $\hat{C}_{21}(f) - \sigma_\nu^2$: 
    \begin{equation}
        \tilde{C}_{42}(f) = \hat{C}_{42}(f)\left(\hat{C}_{21}(f)-\sigma_\nu^2\right)^{-2}.
        \label{eq:c42_estim_f}
    \end{equation}
    
    Note that $C_{21}$ of a signal is the signal power and does not depend on the modulation type. We denote the $u$'th 
    user's signal's $C_{ki}$ by $C_{ki,u}$. If the value depends on the modulation type or channel access method, it is denoted by 
    $C_{ki,u}(\mathcal{M})$.
    
    Since the $\tilde{C}_{42}$ is normalized by signal power, its mean value is independent of any attenuation due to 
    flat-fading \cite{swami_hierarchical_2000}.  
    
    The additivity of the unnormalized $C_{42}$~\cite[Theorem 2.3.1(vi)]{brillinger_time_2001} can be used to express the mean and 
    variance of the estimated $\tilde{C}_{42}$ when $U$ PUs are transmitting simultaneously:
    \begin{align}
        \E\left[\tilde{C}_{42,U}(f)\middle| \mathcal{M} \right] = & \frac{\sum_{u \in U}C_{21,u}^2 
            C_{42,u}(\mathcal{M})}{\left(\sum_{u\in U}C_{21,u}\right)^2} \label{eq:mean_subset}\\
        \var\left[\tilde{C}_{42,U}(f)\middle| \mathcal{M} \right] = & \frac{\sum_{u \in U}C_{21,u}^4 \var\left[\tilde{C}_{42,u}(f)\middle| \mathcal{M}\right]}{\left(\sum_{u\in U}C_{21,u}\right)^4} \notag \\ 
        & + \frac{\sigma_\nu^8\var\left[\tilde{C}_{42,\nu}(f)\right]}{\left(\sum_{u\in U}C_{21,u}\right)^4}.
        \label{eq:var_subset}
    \end{align}
    Expressions to compute $C_{42,u}(\mathcal{M})$ and $\var\left[\tilde{C}_{42,u}(f) \middle| \mathcal{M}\right]$ are
    provided in the Appendix while values for noise-less signals are listed in Table~\ref{tab:cumulants} as computed
        from (\ref{eq:mean_subset}) and (\ref{eq:var_subset}).  In the interest of readability, the left hand side of
    (\ref{eq:mean_subset}) and (\ref{eq:var_subset}) does not explicitly mention the conditional dependence of the mean
    and variance on the signal powers and noise variance.  \footnotetext[1]{Derivation of $\var[\hat{C}_{42}]$ in
    \cite{swami_hierarchical_2000} missed an $O(1/J)$ term.  Details provided in the appendix.}

    \begin{table}
        \caption{Sample mean and variance of $C_{42}$ estimated from $J$ samples for unit power noise-less signals}
        \label{tab:cumulants}
        \centering
        \begin{tabular}{c|c||c|c}
            \hline 
            Constellation & $C_{42}$ & $J\var(\hat{C}_{42})$ from \cite{swami_hierarchical_2000} \textsuperscript{1}& $J\var(\hat{C}_{42})$\tabularnewline
            \hline 
            \hline 
            BPSK & -2.0000 & 36.00 & 0.00\tabularnewline
            \hline 
            \hline 
            PAM(4) & -1.3600 & 34.72 & 10.24\tabularnewline
            \hline 
            PAM(8) & -1.2381 & 32.27 & 9.98\tabularnewline
            \hline 
            PAM(16) & -1.2094 & 31.67 & 9.90\tabularnewline
            \hline 
            PAM(32) & -1.2024 & 31.52 & 9.88\tabularnewline
            \hline 
            PAM(64) & -1.2006 & 31.49 & 9.88\tabularnewline
            \hline 
            PAM($\infty$) & -1.2000 & 31.47 & 9.87\tabularnewline
            \hline 
            \hline 
            PSK($\geq$4) & -1.0000 & 12.00 & 0.00\tabularnewline
            \hline 
            \hline 
            V32 & -0.6900 & 9.70 & 1.42\tabularnewline
            \hline 
            V29 & -0.5816 & 8.75 & 1.77\tabularnewline
            \hline 
            QAM(4,4) & -0.6800 & 9.54 & 1.38\tabularnewline
            \hline 
            QAM(8,8) & -0.6191 & 8.82 & 1.39\tabularnewline
            \hline 
            QAM(16,16) & -0.6047 & 8.65 & 1.39\tabularnewline
            \hline 
            QAM(32,32) & -0.6012 & 8.61 & 1.39\tabularnewline
            \hline 
            QAM($\infty$ ) & -0.6000 & 8.59 & 1.39\tabularnewline
            \hline
            \hline
            BPSK-OFDM & 0 & -- & $\sim$8\tabularnewline
            \hline
            QPSK-OFDM & 0 & -- & $\sim$4\tabularnewline
            \hline
        \end{tabular}
    \end{table}

    \subsection{Proposed Method}

    Since collisions would modify test statistics that depend on the signal structure, we use a 2 stage algorithm that
    classifies between TDMA, CDMA, and OFDMA first and then detect whether there are collisions in the second stage. The
    first stage is a multihypothesis test using the sample mean of $\tilde{C}_{42}(f)$ to classify as TDMA, CDMA, or OFDMA.
    This is an extension of the modulation classification method proposed in~\cite{swami_hierarchical_2000}. The second
    stage proposes a novel binary hypothesis test to detect collisions by thresholding the sample variance of
    $\tilde{C}_{42}(f)$.
    
    We divide the squelched received signal $\{y(n)\}_{n \in \{1,\ldots,JF\}}$ into $F$ frames of $J$ samples each.
    As proposed in~\cite{swami_hierarchical_2000}, The sample mean $W$ of the normalized $C_{42}$ can be used to identify the modulation type of the received signal. We now extend it to identify TDMA, OFDMA, and CDMA. In particular, we classify the received signal as being one of the classes listed in Table~\ref{tab:list_classes}. Let $\tilde{C}_{42, U}(f)$ be the estimated normalized $C_{42}$ if $U \subseteq \mathcal{U}(f)$ indexed PUs were transmitting in frame $f$. Let $\mathcal{U}(f)$ denote the set of PUs transmitting simultaneously in frame $f$. Then, the measured $\tilde{C}_{42}(f)$ can be written as:
    \begin{equation}
        \tilde{C}_{42}(f) = \sum_{U \subseteq \mathcal{U}} 1_{\{U = \mathcal{U}(f)\}} \tilde{C}_{42,U}(f).
        \label{eq:c42f_sum}
    \end{equation}
    
    Since $\tilde{C}_{42}(f)$ has a Gaussian distribution due to the central limit theorem, (\ref{eq:c42f_sum}) implies that $\tilde{C}_{42}(f)$ has a Gaussian mixture distribution with mean and variance given by
    \begin{align}
        & \E\left[\tilde{C}_{42}(f)\middle| \mathcal{M}\right] = \sum_{U \subseteq \mathcal{U}} P\left(U=\mathcal{U}(f)\right) \E\left[\tilde{C}_{42,U}(f)\middle|\mathcal{M}\right] \label{eq:mean_c42f} \\
        &\var\left[\tilde{C}_{42}(f)\middle| \mathcal{M}\right] = \sum_{U \subseteq \mathcal{U}} P\left(U=\mathcal{U}(f)\right) \left\{ \var\left[\tilde{C}_{42,U}(f) \middle| \mathcal{M}\right] \right. \notag \\ 
        & \hspace{4.2em} \left. + \left(\E\left[\tilde{C}_{42,U}(f)\middle| \mathcal{M}\right] - E\left[\tilde{C}_{42}(f)\middle| \mathcal{M}\right] \right)^2 \right\}. \label{eq:var_c42f}
    \end{align}
    
    Further, assuming that the same number, say $K$, of PUs transmit at any time, then (\ref{eq:mean_c42f}) and (\ref{eq:var_c42f}) can be simplified to
    \begin{align}
        \E\left[\tilde{C}_{42}(f)\middle|\mathcal{M}\right] &= C_{42,U'}(\mathcal{M}) \text{ and} \label{eq:stage1_mean}\\
        \var\left[\tilde{C}_{42}(f)\middle|\mathcal{M}\right] &= \sum_{\substack{U \in \mathcal{U}\\|U| = K}} P(U = \mathcal{U}(f)) \var\left[\tilde{C}_{42,U}(f)\middle| \mathcal{M}\right] \label{eq:stage1_var}
    \end{align}
    where $U' \subseteq \mathcal{U}$ and $|U'| = K$. For TDMA, $K = 1$, while $K = N_{\text{total}}$ for CDMA and OFDMA.
    Since this simplification is not possible for contention-based channel access methods, we separate the
    contention detection to the second stage.
    
    The first stage of our algorithm uses (\ref{eq:stage1_mean}), (\ref{eq:stage1_var}), and the sample mean $W$ of $\tilde{C}_{42}(f)$ to find the most likely class $\hat{\mathcal{M}}$ amongst those listed in Table~\ref{tab:list_classes}.
    \begin{equation}
        \hat{\mathcal{M}} = \argmax_{\mathcal{M}' \in \mathfrak{M}} P\left(C_{42} = W | \mathcal{M}' \right)
        \label{eq:MC_ML_estim}
    \end{equation}
    
    For the second stage of our algorithm, detection of contention, we use (\ref{eq:var_c42f}) to note that varying number of simultaneously transmitting PUs increases the sample variance:
    \begin{equation*}
        \var\left[\tilde{C}_{42}(f) \middle| \mathcal{M} \not\in \mathfrak{M}\right] > \var\left[\tilde{C}_{42}(f) \middle| \mathcal{M} \in \mathfrak{M} \right].
    \end{equation*}
    We use this fact to estimate our confidence in the inference $\hat{\mathcal{M}}$. We define $P_{C|T} \in (0, 1)$ as the maximum probability of detecting collisions for non-contention based medium access control protocols. We compute the second moment of $\tilde{C}_{42}(f)$ around the theoretically expected $C_{42}(\hat{\mathcal{M}})$:
    \begin{equation}
        \hat{\varsigma}^2 = \frac{1}{F} \sum_{f=1}^{F} \left( \tilde{C}_{42}(f) - C_{42}\left(\hat{\mathcal{M}}\right) \right)^2.
        \label{eq:svar_c42}
    \end{equation}
    If the correct class has been detected, then $\hat{\varsigma}^2$ is the sample variance of $\tilde{C}_{42}(f)$ and by Cochran's Theorem \cite{cochran_distribution_1934}, the sample variance is distributed as
    \begin{equation}
        \hat{\varsigma}^2 \sim F^{-1} \var\left[\tilde{C}_{42}(f)\middle\vert \mathcal{M}\right]\chi^2_{F}
    \end{equation}
    where $\chi^2_{F}$ is a $\chi^2$ distribution with $F$ degrees. We choose a threshold $\tau$ such that $P_{C|T} > P\left(\varsigma^2 > \tau \middle| \mathcal{M} \not\in \mathcal{H}_C \right)$:
    \begin{equation}
        \tau = F^{-1} \var\left[\tilde{C}_{42}(f) \middle\vert \hat{\mathcal{M}} \right] \chi^{-2}_{F}\left(1-P_{C|T}\right).
        \label{eq:svar_tau}
    \end{equation}
    If $\hat{\varsigma}^2 < \tau$ then we declare the inference $\hat{\mathcal{M}}$ from \eqref{eq:MC_ML_estim} to be correct. If not, then we infer that the channel access method is contention-based.

    \begin{table}[tb]
        \centering
        \caption{List of channel access methods and modulations in $\mathfrak{M}$ in \eqref{eq:MC_ML_estim}.}
        \label{tab:list_classes}
        \begin{tabular}{|c|c|c|}
            \hline
            Class Label & Channel Access Method & Modulation Type \& Levels \\
            \hline
            M1 & \multirow{4}{*}{TDMA} & BPSK \\
            M2 &  & 4/8/16/32/64-PSK \\
            M3-M7 &  & 4/8/16/32/64-PAM \\
            M8-M10 &  & 16/64/256-QAM \\
            \hline
            M11 & \multirow{2}{*}{OFDMA} & 4-QAM\\
            M12 &  & 16-QAM \\
            \hline
            M13 & \multirow{3}{*}{CDMA} & BPSK \\
            M14 & & 4-QAM \\
            M15 & & 16-QAM \\
            \hline
        \end{tabular}
    \end{table}

    \section{Results and Comparisons}
    \label{sec:Results}
    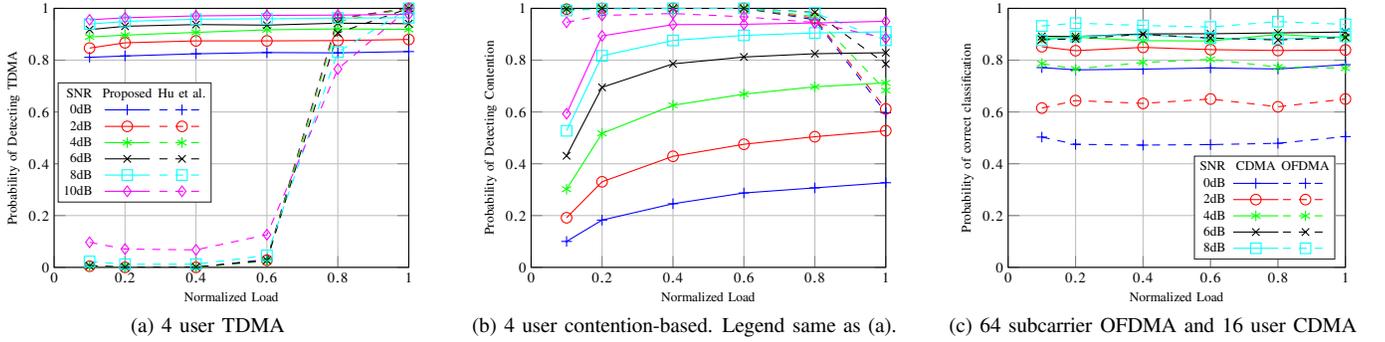
\begin{figure*}[t]
    \centering
    \subfloat[4 user TDMA]{\tiny
        \setlength\figureheight{0.19\textwidth}
        \setlength\figurewidth{0.26\textwidth}
%
\definecolor{mycolor1}{rgb}{0.00000,1.00000,1.00000}%
\definecolor{mycolor2}{rgb}{1.00000,0.00000,1.00000}%
\begin{tikzpicture}

\begin{axis}[%
width=\figurewidth,
height=\figureheight,
at={(0\figurewidth,0\figureheight)},
scale only axis,
xmin=0,
xmax=1,
xlabel={Normalized Load},
xmajorgrids,
ymin=0,
ymax=1,
ylabel={Probability of Detecting TDMA},
ymajorgrids,
axis background/.style={fill=white},
legend columns=2,
legend style={at={(axis cs:0.01,0.25)},anchor=south west,legend plot pos=right,legend cell align=right,align=left,draw=white!15!black,/tikz/every even column/.append style={column sep=-0.2cm}, row sep=-2.0pt,inner sep=0pt,font=\tiny},
]
\addlegendentry{SNR}
\addlegendimage{legend image with text={Proposed}};
\addlegendentry{};
\addlegendimage{legend image with text={Hu et al.}};

\addplot [color=blue,solid,mark=+,mark options={solid}]
  table[row sep=crcr]{%
0.1	0.811071428571429\\
0.2	0.816214285714286\\
0.4	0.824642857142857\\
0.6	0.829357142857143\\
0.8	0.828214285714286\\
1	0.833142857142857\\
};
\addlegendentry{0dB};

\addplot [color=blue,dashed,mark=+,mark options={solid}]
  table[row sep=crcr]{%
0.1	0.00706766917293233\\
0.2	0.000150375939849624\\
0.4	0\\
0.6	0.0309774436090226\\
0.8	0.962180451127819\\
1	1\\
};
\addlegendentry{};

\addplot [color=red,solid,mark=o,mark options={solid}]
  table[row sep=crcr]{%
0.1	0.846642857142857\\
0.2	0.867071428571429\\
0.4	0.874428571428571\\
0.6	0.8745\\
0.8	0.875714285714286\\
1	0.879857142857143\\
};
\addlegendentry{2dB};

\addplot [color=red,dashed,mark=o,mark options={solid}]
  table[row sep=crcr]{%
0.1	0.0043609022556391\\
0.2	0\\
0.4	0\\
0.6	0.0273684210526316\\
0.8	0.959398496240601\\
1	1\\
};
\addlegendentry{};

\addplot [color=green,solid,mark=asterisk,mark options={solid}]
  table[row sep=crcr]{%
0.1	0.889714285714286\\
0.2	0.896642857142857\\
0.4	0.907214285714286\\
0.6	0.917071428571429\\
0.8	0.9205\\
1	0.918714285714286\\
};
\addlegendentry{4dB};

\addplot [color=green,dashed,mark=asterisk,mark options={solid}]
  table[row sep=crcr]{%
0.1	0.00451127819548872\\
0.2	0\\
0.4	7.5187969924812e-05\\
0.6	0.0248872180451128\\
0.8	0.949473684210526\\
1	1\\
};
\addlegendentry{};

\addplot [color=black,solid,mark=x,mark options={solid}]
  table[row sep=crcr]{%
0.1	0.917928571428571\\
0.2	0.930857142857143\\
0.4	0.9375\\
0.6	0.935285714285714\\
0.8	0.942857142857143\\
1	0.940857142857143\\
};
\addlegendentry{6dB};

\addplot [color=black,dashed,mark=x,mark options={solid}]
  table[row sep=crcr]{%
0.1	0.0056390977443609\\
0.2	0.000977443609022556\\
0.4	0.00150375939849624\\
0.6	0.0269924812030075\\
0.8	0.904962406015038\\
1	1\\
};
\addlegendentry{};

\addplot [color=mycolor1,solid,mark=square,mark options={solid}]
  table[row sep=crcr]{%
0.1	0.939214285714286\\
0.2	0.949642857142857\\
0.4	0.958214285714286\\
0.6	0.9595\\
0.8	0.962571428571429\\
1	0.963714285714286\\
};
\addlegendentry{8dB};

\addplot [color=mycolor1,dashed,mark=square,mark options={solid}]
  table[row sep=crcr]{%
0.1	0.0226315789473684\\
0.2	0.0121804511278195\\
0.4	0.012406015037594\\
0.6	0.0462406015037594\\
0.8	0.831353383458647\\
1	1\\
};
\addlegendentry{};

\addplot [color=mycolor2,solid,mark=diamond,mark options={solid}]
  table[row sep=crcr]{%
0.1	0.955642857142857\\
0.2	0.963714285714286\\
0.4	0.970928571428571\\
0.6	0.972928571428571\\
0.8	0.973928571428571\\
1	0.976785714285714\\
};
\addlegendentry{10dB};

\addplot [color=mycolor2,dashed,mark=diamond,mark options={solid}]
  table[row sep=crcr]{%
0.1	0.0966165413533834\\
0.2	0.0706766917293233\\
0.4	0.0675187969924812\\
0.6	0.12609022556391\\
0.8	0.76578947368421\\
1	1\\
};
\addlegendentry{};

\end{axis}
\end{tikzpicture}%
        \label{fig:tdma_detect_4_users_5dB}
    }\hfill
    \subfloat[4 user contention-based. Legend same as (a).]{\tiny
        \setlength\figureheight{0.19\textwidth}
        \setlength\figurewidth{0.26\textwidth}
%
\definecolor{mycolor1}{rgb}{0.00000,1.00000,1.00000}%
\definecolor{mycolor2}{rgb}{1.00000,0.00000,1.00000}%
\begin{tikzpicture}

\begin{axis}[%
width=\figurewidth,
height=\figureheight,
at={(0\figurewidth,0\figureheight)},
scale only axis,
xmin=0,
xmax=1,
xlabel={Normalized Load},
xmajorgrids,
ymin=0,
ymax=1,
ylabel={Probability of Detecting Contention},
ymajorgrids,
axis background/.style={fill=white},
legend columns=2,
legend pos={south east},
legend style={legend plot pos=right,legend cell align=left,align=left,draw=white!15!black,/tikz/every even column/.append style={column sep=-0.2cm}, font=\tiny},
ignore legend
]
\addlegendentry{SNR}
\addlegendimage{legend image with text={Proposed}};
\addlegendentry{};
\addlegendimage{legend image with text={Hu et al.}};

\addplot [color=blue,solid,mark=+,mark options={solid}]
  table[row sep=crcr]{%
0.1	0.100142857142857\\
0.2	0.1825\\
0.4	0.245642857142857\\
0.6	0.287214285714286\\
0.8	0.306928571428571\\
1	0.326785714285714\\
};
\addlegendentry{0dB};

\addplot [color=blue,dashed,mark=+,mark options={solid}]
  table[row sep=crcr]{%
0.1	0.996842105263158\\
0.2	1\\
0.4	1\\
0.6	1\\
0.8	0.955789473684211\\
1	0.596842105263158\\
};
\addlegendentry{};

\addplot [color=red,solid,mark=o,mark options={solid}]
  table[row sep=crcr]{%
0.1	0.191428571428571\\
0.2	0.330357142857143\\
0.4	0.428785714285714\\
0.6	0.475214285714286\\
0.8	0.504428571428571\\
1	0.5275\\
};
\addlegendentry{2dB};

\addplot [color=red,dashed,mark=o,mark options={solid}]
  table[row sep=crcr]{%
0.1	0.996842105263158\\
0.2	1\\
0.4	1\\
0.6	1\\
0.8	0.962105263157895\\
1	0.611578947368421\\
};
\addlegendentry{};

\addplot [color=green,solid,mark=asterisk,mark options={solid}]
  table[row sep=crcr]{%
0.1	0.301857142857143\\
0.2	0.517142857142857\\
0.4	0.625785714285714\\
0.6	0.668785714285714\\
0.8	0.697142857142857\\
1	0.712428571428571\\
};
\addlegendentry{4dB};

\addplot [color=green,dashed,mark=asterisk,mark options={solid}]
  table[row sep=crcr]{%
0.1	1\\
0.2	1\\
0.4	1\\
0.6	1\\
0.8	0.968421052631579\\
1	0.683157894736842\\
};
\addlegendentry{};

\addplot [color=black,solid,mark=x,mark options={solid}]
  table[row sep=crcr]{%
0.1	0.430857142857143\\
0.2	0.695142857142857\\
0.4	0.785714285714286\\
0.6	0.812214285714286\\
0.8	0.824857142857143\\
1	0.828785714285714\\
};
\addlegendentry{6dB};

\addplot [color=black,dashed,mark=x,mark options={solid}]
  table[row sep=crcr]{%
0.1	0.997894736842105\\
0.2	1\\
0.4	1\\
0.6	1\\
0.8	0.984210526315789\\
1	0.785263157894737\\
};
\addlegendentry{};

\addplot [color=mycolor1,solid,mark=square,mark options={solid}]
  table[row sep=crcr]{%
0.1	0.527285714285714\\
0.2	0.816785714285714\\
0.4	0.876214285714286\\
0.6	0.895714285714286\\
0.8	0.904928571428571\\
1	0.909357142857143\\
};
\addlegendentry{8dB};

\addplot [color=mycolor1,dashed,mark=square,mark options={solid}]
  table[row sep=crcr]{%
0.1	0.991578947368421\\
0.2	0.997894736842105\\
0.4	1\\
0.6	1\\
0.8	0.982105263157895\\
1	0.876842105263158\\
};
\addlegendentry{};

\addplot [color=mycolor2,solid,mark=diamond,mark options={solid}]
  table[row sep=crcr]{%
0.1	0.593285714285714\\
0.2	0.893857142857143\\
0.4	0.937142857142857\\
0.6	0.9385\\
0.8	0.943714285714286\\
1	0.95\\
};
\addlegendentry{10dB};

\addplot [color=mycolor2,dashed,mark=diamond,mark options={solid}]
  table[row sep=crcr]{%
0.1	0.946315789473684\\
0.2	0.972631578947368\\
0.4	0.98\\
0.6	0.967368421052632\\
0.8	0.945263157894737\\
1	0.884210526315789\\
};
\addlegendentry{};

\end{axis}
\end{tikzpicture}%
        \label{fig:contention_detect_4_users_5dB}
	}\hfill
    \subfloat[64 subcarrier OFDMA and 16 user CDMA]{\tiny
        \setlength\figureheight{0.19\textwidth}
        \setlength\figurewidth{0.26\textwidth}
%
\definecolor{mycolor1}{rgb}{0.00000,1.00000,1.00000}%
\begin{tikzpicture}

\begin{axis}[%
width=0.951\figurewidth,
height=\figureheight,
at={(0\figurewidth,0\figureheight)},
scale only axis,
xmin=0,
xmax=1,
xlabel={Normalized Load},
xmajorgrids,
ymin=0,
ymax=1,
ylabel={Probability of correct classification},
ymajorgrids,
axis background/.style={fill=white},
transpose legend,
legend columns=6,
legend style={legend pos=south east,legend plot pos=right,legend cell align=right,align=left,draw=white!15!black,/tikz/every even column/.append style={column sep=-0.2cm}, row sep=-2.0pt,inner sep=0pt,font=\tiny},
]
\addlegendentry{SNR}
\addlegendimage{legend image with text={CDMA}};

\addplot [color=blue,solid,mark=+,mark options={solid}]
  table[row sep=crcr]{%
0.1	0.771333333333333\\
0.2	0.762666666666667\\
0.4	0.765333333333333\\
0.6	0.77\\
0.8	0.766\\
1	0.782666666666667\\
};
\addlegendentry{0dB};

\addplot [color=red,solid,mark=o,mark options={solid}]
  table[row sep=crcr]{%
0.1	0.852\\
0.2	0.836666666666667\\
0.4	0.849333333333333\\
0.6	0.840666666666667\\
0.8	0.837333333333333\\
1	0.839333333333333\\
};
\addlegendentry{2dB};

\addplot [color=green,solid,mark=asterisk,mark options={solid}]
  table[row sep=crcr]{%
0.1	0.884666666666667\\
0.2	0.888666666666667\\
0.4	0.875333333333333\\
0.6	0.874\\
0.8	0.899333333333333\\
1	0.886\\
};
\addlegendentry{4dB};

\addplot [color=black,solid,mark=x,mark options={solid}]
  table[row sep=crcr]{%
0.1	0.892\\
0.2	0.891333333333333\\
0.4	0.901333333333333\\
0.6	0.902\\
0.8	0.905333333333333\\
1	0.908666666666667\\
};
\addlegendentry{6dB};

\addplot [color=mycolor1,solid,mark=square,mark options={solid}]
  table[row sep=crcr]{%
0.1	0.876666666666667\\
0.2	0.891333333333333\\
0.4	0.898\\
0.6	0.884666666666667\\
0.8	0.882\\
1	0.891333333333333\\
};
\addlegendentry{8dB};

\addlegendentry{};
\addlegendimage{legend image with text={OFDMA}};

\addplot [color=blue,dashed,mark=+,mark options={solid}]
  table[row sep=crcr]{%
0.1	0.503\\
0.2	0.475\\
0.4	0.472\\
0.6	0.474\\
0.8	0.479\\
1	0.505\\
};
\addlegendentry{};

\addplot [color=red,dashed,mark=o,mark options={solid}]
  table[row sep=crcr]{%
0.1	0.615\\
0.2	0.644\\
0.4	0.633\\
0.6	0.65\\
0.8	0.62\\
1	0.65\\
};
\addlegendentry{};

\addplot [color=green,dashed,mark=asterisk,mark options={solid}]
  table[row sep=crcr]{%
0.1	0.787\\
0.2	0.766\\
0.4	0.791\\
0.6	0.803\\
0.8	0.773\\
1	0.769\\
};
\addlegendentry{};

\addplot [color=black,dashed,mark=x,mark options={solid}]
  table[row sep=crcr]{%
0.1	0.879\\
0.2	0.882\\
0.4	0.9\\
0.6	0.885\\
0.8	0.878\\
1	0.888\\
};
\addlegendentry{};

\addplot [color=mycolor1,dashed,mark=square,mark options={solid}]
  table[row sep=crcr]{%
0.1	0.932\\
0.2	0.943\\
0.4	0.934\\
0.6	0.928\\
0.8	0.949\\
1	0.938\\
};
\addlegendentry{};

\end{axis}
\end{tikzpicture}%
        \label{fig:cdma_ofdma}
    }
    \caption{Probability of correctly identifying the channel access method for TDMA, OFDMA, and CDMA and comparison with~\cite{hu_mac_2014} for detection of TDMA.}
    \label{fig:non_contention_5dB}
    \end{figure*}

    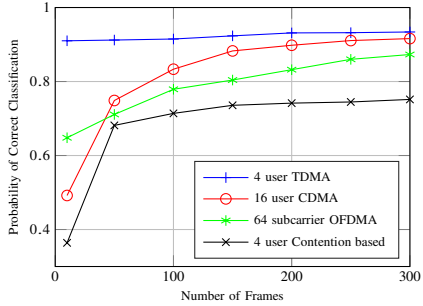
\begin{figure}
        \centering
        {\tiny
        \setlength\figureheight{0.19\textwidth}
        \setlength\figurewidth{0.26\textwidth}
%
\definecolor{mycolor1}{rgb}{0.00000,1.00000,1.00000}%
\definecolor{mycolor2}{rgb}{1.00000,0.00000,1.00000}%
\begin{tikzpicture}

\begin{axis}[%
width=\figurewidth,
height=\figureheight,
at={(0\figurewidth,0\figureheight)},
scale only axis,
xmin=0,
xmax=300,
xlabel={Number of Frames},
xmajorgrids,
ymin=0.3,
ymax=1,
ylabel={Probability of Correct Classification},
ymajorgrids,
axis background/.style={fill=white},
legend pos={south east},
legend style={legend cell align=left,align=left,draw=white!15!black,font=\tiny},
]

\addplot [color=blue,solid,mark=+,mark options={solid}]
  table[row sep=crcr]{%
 10 0.9101\\
 50 0.9121\\
100 0.9149\\
150 0.9234\\
200 0.9314\\
250 0.9320\\
300 0.9339\\
};
\addlegendentry{4 user TDMA};

\addplot [color=red,solid,mark=o,mark options={solid}]
  table[row sep=crcr]{%
 10 0.4920\\
 50 0.7487\\
100 0.8333\\
150 0.8827\\
200 0.8980\\
250 0.9107\\
300 0.9160\\
};
\addlegendentry{16 user CDMA};

\addplot [color=green,solid,mark=asterisk,mark options={solid}]
  table[row sep=crcr]{%
 10 0.6480\\
 50 0.7110\\
100 0.7790\\
150 0.8040\\
200 0.8320\\
250 0.8600\\
300 0.8730\\
};
\addlegendentry{64 subcarrier OFDMA};

\addplot [color=black,solid,mark=x,mark options={solid}]
  table[row sep=crcr]{%
 10 0.3640\\
 50 0.6814\\
100 0.7139\\
150 0.7357\\
200 0.7416\\
250 0.7447\\
300 0.7516\\
};
\addlegendentry{4 user Contention based};

\end{axis}
\end{tikzpicture}%
        }
        \caption{Probability of correctly identifying channel access methods as number of frames are increased. System:
        $J = 500$, 5dB SNR at 0.5 normalized load.}
        \label{fig:varying_nframes}
    \end{figure}

    The theoretical distributions of both the proposed test statistics have been described in the previous section.
    Since our proposed algorithm consists of a multi-hypothesis test, it is not possible to derive a closed form
    expression for the classification accuracy. Hence, in this section, we use simulations to study the performance of
    our proposed channel access method classification algorithm. 

    \subsection{Simulation System}
    Consider $N$ PUs communicating by a TDMA, CDMA, OFDMA, or contention-based channel access method. The signals
    transmitted by these users are as described in Section~\ref{sec:ProblemFormulation}.  We assume a Rayleigh flat
    fading channel between the PUs and our sensor node. The received SNR of each individual user is exponentially
    distributed \cite{proakis_digital_2007} and we vary the average SNR. The offered load by the system is varied from
    0.1 to 1. Our metric is the probability of correctly classifying channel access methods. We have chosen the
    parameter $P_{\text{C}|\text{T}}$ as 0.05. We present results averaged over the classes listed in
    Table~\ref{tab:list_classes}.

    We also implemented the SVM-based classifier proposed by Hu et al. in~\cite{hu_mac_2014} to distinguish between TDMA
    and slotted ALOHA. The SVM is trained using features of received energy, idle time, and busy time of the channel. In
    order to simulate a blind scenario, we trained the SVM with about 60,000 realizations consisting of 50 realizations
    from each modulation type, SNR, number of users, and traffic load.

    \subsection{Effect of Traffic Load}
    \label{sec:cam_sims}

    The traffic load does not affect the classification of TDMA, CDMA, and OFDMA because we squelch the input signal.
    Fig.~\ref{fig:non_contention_5dB} shows this classification accuracy as a function of the normalized load. However,
    the number of collisions in a contention-based method increase with load and causes the classification accuracy of
    contention-based channel access method to increase with normalized load.

    SNR affects the variance of $\tilde{C}_{42}(f)$ for all classes but its mean is affected only for contention-based
    channel access methods. Therefore, low SNR affects the classification of contention-based methods more than that of
    the other classes. Furthermore, in case a few ``hidden'' PUs have very low SNRs due to, say, distance
    or deep fading channels, the classification accuracy will reduce for all classes and the detection of contention
    will suffer the most. However, if the remaining PUs have higher SNR, then the classification accuracy will
    increase.

    Further, Figs.~\subref*{fig:tdma_detect_4_users_5dB} and~\subref*{fig:contention_detect_4_users_5dB} compares with
    the SVM based classifier proposed in~\cite{hu_mac_2014} which considers only TDMA and contention-based schemes.
    Since~\cite{hu_mac_2014} separates only two classes, note that our proposed algorithm has significantly
     higher combined classification accuracy. The SVM proposed in~\cite{hu_mac_2014} tends to classify all signals as
    contention-based at low loads possibly because channel idle and busy times depend more on the load than contention
    for the channel. However, further study is required to improve our algorithm's classification of contention-based
    schemes at low SNRs.

    \subsection{Number of Frames}

    Fig.~\ref{fig:varying_nframes} shows the classification accuracy as the number of frames $F$ is increased.
    Increasing $F$ reduces the sample variance of $\tilde{C}_{42}$ which increases the probability of correctly
    classifying TDMA, CDMA, and OFDMA. Increasing $F$ also increases the probability of observing a collision in
    contention-based methods. So, the probability of correctly classifying contention-based methods increases. For
    comparison, note that Fig.~\ref{fig:non_contention_5dB} shows results for $F = 200$.

    \subsection{Computational Complexity}
    \label{sec:computational_complexity}

    Computing cumulant estimates from $F$ frames of $J$ samples each requires $O(FJ)$ operations. Computing the
    statistics for each class in $\mathcal{M}$ requires $O(|\mathcal{M}|)$ operations. Hence, the first stage of the
    algorithm requires $O(FJ)$ operations.  By reusing values computed in the first stage, computing $\hat{\varsigma}^2$
    and $\tau$ for the second stage requires $O(F)$ operations.
    Thus, our algorithm requires $O(FJ)$ computations dominated by the first stage.

    \section{Conclusion}
    \label{sec:Conclusion}
    In this article, we have presented a new algorithm to identify the channel access method utilized by a primary network. Our methods are not restricted to any specific standards. We extended a cumulant-based modulation type classification technique to differentiate between OFDMA, CDMA, and TDMA. We proposed a novel collision detection method using the sample variance of the cumulant estimator and, thus, identify contention-based channel access methods such as CSMA. These test statistics were chosen to make our methods robust to channel fading and size of the PU network.

\bibliographystyle{IEEEtran}
\bibliography{IEEEabrv,references_v2}

    \appendix[Statistics of Sample Estimates of Cumulants]
    \label{sec:appendix}

    Consider $J$ samples of a complex signal $r(n)$ as given by \eqref{eq:sig_model}. Define $y(n) = r(n) - \frac{1}{J}\sum_{n=1}^J r(n)$ so as to obtain a zero-mean form of the received signal. We wish to derive the mean and variance of the estimator $\tilde{C}_{42,y}$ where the subscript $y$ indicates that the cumulant is computed from the signal $y(n)$.
    
    We begin by studying the unbiased estimator for the (unnormalized) $\hat{C}_{42,y}$ cumulant:
    \begin{equation}
        \hat{C}_{42,y}=M_{42,y}-|\hat{C}_{20,y}|^{2}-2\hat{C}_{21,y}^{2}.
    \end{equation} 
    Since these terms are correlated, we have
    \begin{align}
        \var[\hat{C}_{42,y}] = &\var[\hat{M}_{42,y}]+\var[|\hat{M}_{20,y}|^{2}]+4\var[\hat{M}_{21,y}^2] \notag \\
        &-2\cov[\hat{M}_{42,y},|\hat{M}_{20,y}|^2] - 4\cov[\hat{M}_{42,y},\hat{M}_{21,y}^2] \notag \\
        &+ 4\cov[|\hat{M}_{20,y}|^2,\hat{M}_{21,y}^2].
        \label{eqn:varc42}
    \end{align}
    From the Appendix in \cite{swami_hierarchical_2000} we have 
    \begin{equation}
        \var[\hat{M}_{42,y}]=\frac{1}{J}(M_{84,y}-M_{42,y}^{2})
        \label{eqn:varm21}
    \end{equation}
    The asymptotic analysis of both $\var[\hat{M}_{21,y}^2]$ and $\cov[\hat{M}_{42,y},\hat{M}_{21,y}^2]$ as derived in \cite{swami_hierarchical_2000} are incorrect since they fail to take into account some $O(1/J)$ terms as shown in the following derivation. Using the definition $\alpha \triangleq M_{42,y}-M_{21,y}^2$,
    \begin{align}
        \var[\hat{M}_{21,y}^{2}]=&\frac{(J-1)(J-2)(J-3)}{J^{3}}M_{21,y}^{4} \notag \\
        &+\frac{6(J-1)(J-2)}{J^{3}}M_{21,y}^{2}M_{42,y}^{2}\nonumber\\
        &-\left(M_{21,y}^{2}+\frac{\alpha}{J}\right)^{2}+O(1/J^{2})\nonumber\\
        =&\left(1-\frac{6}{J}\right)M_{21,y}^{4}+\frac{6}{J}M_{21,y}^{2}M_{42,y} \notag \\
        &-\left(M_{21,y}^{2}+\frac{\alpha}{J}\right)^{2}+O(1/J^{2})\nonumber\\
        \approx&\frac{4}{J}M_{21,y}^2(M_{42,y}-M_{21,y}^2)
    \end{align}
    The error is in failing to take into account the $-\frac{6}{J}M_{21,y}^4$ part of the first term. A similar derivation  also gives the corrected expression
    \begin{equation}
        \cov[\hat{M}_{42,y},\hat{M}_{21,y}^{2}]=\frac{2}{J}M_{21,y}(M_{63,y}-M_{42,y}M_{21,y})
    \end{equation}
    Following similar derivations we can find the rest of the terms in (\ref{eqn:varc42}) as
    \begin{align}
        \var[|\hat{M}_{20,y}|^2]=&\frac{2}{J}M_{42,y}|M_{20,y}|^{2}-\frac{4}{J}|M_{20,y}|^{4} \notag \\
        & +\frac{2}{J}\re\{M_{40,y}M_{20,y}^{*2}\}\\
        \cov[\hat{M}_{42,y},|\hat{M}_{20,y}|^{2}]=&\frac{2}{J}\re\{M_{62,y}M_{20,y}^{*}\} \notag \\
        & -\frac{2}{J}M_{42,y}|{M}_{20,y}|^{2}\\
        \cov[|\hat{M}_{20,y}|^{2},\hat{M}_{21,y}^{2}]=&\frac{4}{J}M_{21,y}\re\{M_{41,y}M_{20,y}^{*}\} \notag \\
        & -\frac{4}{J}M_{21,y}^{2}|{M}_{20,y}|^{2}
        \label{eqn:covm20m21}
    \end{align}
    Substituting (\ref{eqn:varm21})--(\ref{eqn:covm20m21}) into (\ref{eqn:varc42}) we find the general expression for the asymptotic variance of the $C_{42,y}$ estimate as follows:
    \begin{align}
        J & \var[\hat{C}_{42,y}]\approx M_{84,y}-M_{42,y}^{2} \notag \\
        & + 8M_{21,y}\left[2M_{21,y}(M_{42,y}-M_{21,y}^{2}-|M_{20,y}|^{2}) \right. \notag \\
        & \hspace{5em} \left. +2\re\{M_{41,y}M_{20,y}^{*}\} - M_{63,y}+M_{21,y}M_{42,y}\right] \notag \\
        & + 2 \re\{M_{20,y}^{*}(M_{40,y}M_{20,y}^{*}-2M_{62,y})\}\nonumber \\
        &+2|M_{20,y}|^{2}(3M_{42,y}-2|M_{20,y}|^{2}).
        \label{eqn:generalvar}
    \end{align}

    Now, $y(n)$ is a noisy signal. We will rewrite \eqref{eqn:generalvar} in terms of cumulants so that we can quantify the effect of noise using the additive property of cumulants.
    The following moment--cumulant equivalence relations are easy to derive:
    \begin{align}
        \label{eq:moment_to_cumulant_1}
        M_{84,y}=& C_{84,y} + 16C_{63,y}C_{21,y} + 12\re\{C_{64,y}C_{20,y}\} \notag \\
        &+ 72C_{21,y}^2C_{42,y} + 18C_{42,y}^2 + 16|C_{41,y}|^2 \notag \\
        &+ |C_{40,y}|^2 + 6\re\{C_{40,y}^*C_{20,y}^2\} \notag \\
        &+ 96\re\{C_{41,y}^*C_{20,y}\}C_{21,y} + 36|C_{20,y}|^2C_{42,y} \notag \\
        &+ 72|C_{20,y}|^2C_{21,y}^2 + 24C_{21,y}^4 + 9|C_{20,y}|^4 \\
        M_{63,y} =& C_{63,y} + 6\re[C_{20,y}C_{43,y}] + 9|C_{20,y}|^2 C_{21,y} + 6C_{21,y}^3 \notag \\
        & +9C_{21,y}C_{42,y} \\
        M_{42,y} =& C_{42,y}+|C_{20,y}|^2 + 2C_{21,y}^2\notag \\
        M_{40,y} =& C_{40,y} + 3C_{20,y}^2 \notag \\
        M_{21,y} =& C_{21,y}.
        \label{eq:moment_to_cumulant_2}
    \end{align}
    Gaussian noise has all the relevant cumulants zero except for $C_{21,\nu} = \sigma_\nu^2$. By slight abuse of notation, let $C_{ki,x}$ be the $C_{ki}$ cumulant of the noiseless signal component of $y(n)$. Then, except for $C_{21,y}$, we can rewrite all the relevant cumulants as $C_{ki,y}=C_{ki,x}$. $C_{21,y}$ can be rewritten as $C_{21,y}=C_{21,x}+\sigma_\nu^2$.
    Using these relations and \eqref{eq:moment_to_cumulant_1}-\eqref{eq:moment_to_cumulant_2},
    we can rewrite \eqref{eqn:generalvar} in terms of cumulants.
    
    \subsection{Single User Signals}
    If it is known that the received signal $r(n)$ consists of a single user's signal, i.e., no collisions have occurred and it is not a CDMA or OFDMA signal, then we can use the modulation type $M$ of the signal to describe $\var\left[\tilde{C}_{42,y}\right]$. We do this by assuming that the normalizing factor $\left(\hat{C}_{21,y}-\sigma_\nu^2\right)^2$ is perfectly estimated. Then, after normalization, $C_{ki,x}$ are replaced by $C_{ki}(M)$ where $C_{ki}(M)$ is the $C_{ki}$ cumulant of a unit power signal having modulation $M$. After normalization, $C_{21,y}$ would be replaced by $C_{21,y}/(C_{21,y}-\sigma_\nu^2)$.
    Using these relations and \eqref{eq:moment_to_cumulant_1}-\eqref{eq:moment_to_cumulant_2},
    we can rewrite \eqref{eqn:generalvar} in terms of cumulants for signals modulated by real constellations as:
    \begin{align}
        & J\var[\tilde{C}_{42,y}] = C_{84}(M) + 4C_{63}(M)\left[\frac{C_{21,y}}{C_{21,y}-\sigma_\nu^2}\right] \notag \\
        &+ 12\re\{C_{62}^*(M)C_{20}(M)\} + 17C_{42}(M)^2 \notag \\
        &- 8\left[\frac{C_{21,y}}{C_{21,y}-\sigma_\nu^2}\right]^2C_{42}(M) + 34|C_{20}(M)|^2C_{42}(M) \notag \\
        &+ 16|C_{41}(M)|^2 + 24\re\{C_{41}(M)^*C_{20}(M)\}\left[\frac{C_{21,y}}{C_{21,y}-\sigma_\nu^2}\right] \notag \\
        & + |C_{40}(M)|^2 + 6\re\{C_{40}(M)^*C_{20}(M)^2\} \notag \\
        & + 24\left[\frac{C_{21,y}}{C_{21,y}-\sigma_\nu^2}\right]^4
        \label{eq:cum_var_awgn_real}
    \end{align}
    where we use the fact that real constellations have all real moments, i.e., $M_{20}=M_{21}$, $M_{40}=M_{41}=M_{42}$, and $M_{62}=M_{63}$.
    Similarly, for signals modulated by constellations having four-fold symmetry, such as QAM, we use $C_{20}=0=C_{21}$ to get
    \begin{align}
        J & \var[\tilde{C}_{42,y}] = C_{84}(M) +|C_{40}(M)|^{2} \notag \\
        & + 8\left[\frac{C_{21,y}}{C_{21,y}-\sigma_\nu^2}\right]C_{63}(M) + 20\left[\frac{C_{21,y}}{C_{21,y}-\sigma_\nu^2}\right]^2C_{42}(M) \notag \\
        & + 4\left[\frac{C_{21,y}}{C_{21,y}-\sigma_\nu^2}\right]^{4} + 17C_{42}(M)^{2} \label{eq:cum_var_awgn_fourfold}
    \end{align}
    This is the corrected form of \cite[Eqns. 13]{swami_hierarchical_2000}.
    
    Table~\ref{tab:cumulants} lists the statistics of the $C_{42}$ estimation for unit power noise-less signals having different modulation types.
    
    \subsection{Statistics for OFDMA Signals}
    \label{sec:c42_stat_ofdma}
    The use of fourth order cumulants for distinguishing OFDM signals from single carrier signals is proposed and analyzed in \cite{shi_fourth_2008}. The moments for OFDM are found to be
    \begin{equation*}
        M_{84}\approx24, \quad M_{63}\approx 6, \quad M_{40}\approx 0, \quad M_{42}\approx2, \quad M_{21}=1.
    \end{equation*}
    Using these moments in the corrected expression in (\ref{eqn:generalvar}) gives that $J \var[\hat{C}_{42}] \approx 4$ which coincidentally also matches the result derived in \cite{shi_fourth_2008} from the incorrect expression of \cite{swami_hierarchical_2000}. 
    Note however that this result
    only applies for OFDM with subcarrier modulations that satisfy the fourfold symmetry such as QPSK. Using these moments the variance of the fourth order cumulant for a noisy OFDM signal can be found using \eqref{eq:cum_var_awgn_fourfold}.
    
    \subsection{Statistics for CDMA Signals} \label{sec:c42_stat_CDMA} %
    A CDMA signal which uses BPSK chips can be viewed as the sum of $N_{\text{total}}$ BPSK signals given as
    \begin{equation*}
        x(n) = \sum_{i=1}^{N_{\text{total}}} s_i(n)
    \end{equation*}
    where $s_i(n)$ is a BPSK formed by spreading the data to be transmitted with the particular code assigned to that user and is given by \eqref{eq:cdma_sig_model}. Thus we can find the mean of the normalized fourth-order cumulant of a noiseless CDMA signal as $E[\hat{C}_{42}]=-2/N_{\text{total}}$ where we invoke the additivity property. In effect, the mean normalized fourth order cumulant approaches 0 as the number of users increases.
    
    As for the variance  of $\hat{C}_{42}$ we can use the general expression in \eqref{eqn:generalvar} once the moments are found. Due to the blind nature of our classification problem, we do not have knowledge of the true spreading code used by each user. As a result, the correlations from one chip to another within the same symbol period cannot be known. However, unlike the single carrier signals presented in this appendix, these correlations are clearly non-zero and are dependent on the codes used. To circumvent this issue we will assume that such correlations are negligible. Note that with this assumption we are treating CDMA signals to be similar to a sum of BPSK signals in which symbols from different symbol periods and different users are regarded as i.i.d. This is clearly an approximation, but we have found through simulations that the discrepancy is negligible in practice.
    
    With this assumption we can proceed to derive the moments of a sum of $N_{\text{total}}$ BPSK signals to be
    \begin{align*}
        M_{42}&=\binom{4}{2}\binom{N_{\text{total}}}{2}\frac{1}{N_{\text{total}}^2} + \frac{1}{N_{\text{total}}}\\
        M_{63}&=\binom{6}{2}\binom{4}{2}\binom{N_{\text{total}}}{3}\frac{1}{N_{\text{total}}^3} + \binom{6}{4}\binom{N_{\text{total}}}{2}\frac{2}{N_{\text{total}}^3} \\
        &\qquad + \frac{1}{N_{\text{total}}^2}\\
        M_{84}&=\binom{8}{2}\binom{6}{2}\binom{4}{2}\binom{N_{\text{total}}}{4}\frac{1}{N_{\text{total}}^4} + \binom{8}{4}\binom{4}{2}\binom{N_{\text{total}}}{3}\frac{3}{J^4}\\
        &\qquad + \left(\binom{8}{6}+\frac{1}{2}\binom{8}{4}\right)\binom{N_{\text{total}}}{2}\frac{2}{N_{\text{total}}^4} + \frac{1}{N_{\text{total}}^3}
    \end{align*}
    The variance of $\tilde{C}_{42}$ of the noisy signals can then be found through \eqref{eq:moment_to_cumulant_2}.

\end{document}